\documentclass[a4paper,english,tightenlines,superscriptaddress,twocolumn,prb,showpacs]{revtex4}
\usepackage{graphicx}
\usepackage{bm}
\usepackage{amsmath}

\begin{document}

\title{Understanding the $\bm{\mu}$SR spectra 
of MnSi without magnetic polarons}
\author{A. Amato}
\email[corresponding author:~]{alex.amato@psi.ch}
\affiliation{Laboratory for Muon Spin Spectroscopy, Paul Scherrer Institut, 5232 Villigen PSI, Switzerland}
\author{P. Dalmas de R\'eotier}
\affiliation{Univ.\ Grenoble Alpes, INAC-SPSMS, F-38000 Grenoble, France}
\affiliation{CEA, INAC-SPSMS, F-38000 Grenoble, France}
\author{D. Andreica}
\affiliation{Faculty of Physics, Babes-Bolyai University, 400084 Cluj-Napoca, Romania}
\author{A.~Yaouanc}
\affiliation{Univ.\ Grenoble Alpes, INAC-SPSMS, F-38000 Grenoble, France}
\affiliation{CEA, INAC-SPSMS, F-38000 Grenoble, France}
\author{A.~Suter}
\affiliation{Laboratory for Muon Spin Spectroscopy, Paul Scherrer Institut, 5232 Villigen PSI, Switzerland}
\author{G.~Lapertot}
\affiliation{Univ.\ Grenoble Alpes, INAC-SPSMS, F-38000 Grenoble, France}
\affiliation{CEA, INAC-SPSMS, F-38000 Grenoble, France}
\author{I.M.~Pop}
\affiliation{Faculty of Physics, Babes-Bolyai University, 400084 Cluj-Napoca, Romania}
\affiliation{Yale University, Applied Physics Department, New Haven, CT 06520-8284, USA}
\author{E. Morenzoni}
\affiliation{Laboratory for Muon Spin Spectroscopy, Paul Scherrer Institut, 5232 Villigen PSI, Switzerland}
\author{P. Bonf\`a}
\affiliation{Dipartimento di Fisica e Scienze della Terra and Unita CNISM di Parma, Universit\`a di Parma, 43124 Parma, Italy}
\author{F. Bernardini}
\affiliation{CNR-IOM-Cagliari and Dipartimento di Fisica, Universita di Cagliari, 09042 Monserrato, Italy}
\author{R. De Renzi}
\affiliation{Dipartimento di Fisica e Scienze della Terra and Unita CNISM di Parma, Universit\`a di Parma, 43124 Parma, Italy}
\begin{abstract} 
Transverse-field muon-spin rotation ($\mu$SR) experiments were performed on a single crystal sample of the non-centrosymmetric system MnSi. The observed angular dependence of the muon precession frequencies 
matches perfectly the one of the Mn-dipolar fields acting on the muons stopping at a 
4$a$ position of the crystallographic structure. The data provide a precise determination of the magnetic dipolar tensor. 
In addition, we have calculated the shape of the field distribution expected below the magnetic transition temperature $T_C$ at the 4$a$ muon-site when no external magnetic field is applied. We show that this field distribution is consistent with the one reported by zero-field $\mu$SR studies.  Finally, we present {\it ab initio} calculations based on the density-functional theory which 
confirm the position of the muon stopping site inferred from transverse-field $\mu$SR.
In view of the presented evidence we conclude that the $\mu$SR response of MnSi can be perfectly and fully understood without invoking a hypothetical magnetic polaron state.
\end{abstract} 
\maketitle
\section{Introduction}
In recent years a growing interest has been focused on crystalline systems lacking inversion symmetry (see for example 
Ref.~\onlinecite{NCS}). A direct consequence of non-centrosymmetry is the occurrence of a spin-orbit Dzyaloshinsky-Moriya (DM) interaction \cite{DM1,DM2}. The DM interaction is thought to play an important role in multiferroic systems and may be at play in the occurrence of non-trivial magnetic structures. On the other hand, non-centrosymmetric systems presenting superconductivity have attracted special attention as the absence of parity symmetry considerably restricts the possible superconducting states (i.e. a pure triplet pairing is in principle not allowed without inversion symmetry) \cite{Sigrist}.

MnSi is probably the non-centrosymmetric system which concentrates most of the research focus. It was the first material found to exhibit a homochiral spin spiral structure below $T_C \simeq 29.5$~K \cite{Ishikawa_1,Ishikawa_2,Ishida}. This spin spiral adopts a very long wavelength of about 18~nm. In addition to an external pressure destroying the magnetic state ($p_c \simeq 1.5$~GPa \cite{Thompson}), a modest magnetic field of 6~kG is found to overcome the DM interaction leading to a spin-aligned state \cite{Ishikawa}. Moreover a skyrmion phase was recently identified near $T_C$ in the magnetic phase diagram \cite{Muhlbauer}. 
Contrary to what observed in a number of strongly correlated electron systems (see for example Ref.~\onlinecite{Goll}), MnSi does not display superconductivity at $p_c$. It was invoked \cite{Saxena} that the absence of a superconducting state for MnSi under pressure has to be traced back to the absence of parity operator for the wavefunctions.

In view of all these properties, MnSi has been investigated several times by $\mu$SR in zero or externally applied pressures (see for example Refs.~\onlinecite{Hayano,Kadono,Yaouanc,Uemura,Andreica}). Two main controversies occurred from these studies. The first one regards the behavior of the magnetic state near the critical pressure $p_c$. First measurements performed by Uemura {\it et al.} \cite{Uemura} seemed to show a phase separation between magnetic and paramagnetic regions whereas measurements performed by Andreica {\it et al.} \cite{Andreica}, extending at lower temperatures, showed a complete absence of phase separation. The second controversy regards the muon stopping site(s) and the occurrence or not of magnetic polaron. The large majority of the studies associates the two spontaneous frequencies seen below $T_C$ to two magnetically inequivalent sites. The same is deduced from transverse-field (TF) measurements performed at low fields \cite{Takigawa}, even though the conclusions obtained from this latter study are blurred by the assumption of a vanishing dipolar tensor. 
On the other side, Storchak {\it et al.} \cite{Storchak} relate the observation of two frequencies in TF data to a coupled $\mu^+e^-$ system (i.e. Mu state). In this scenario, the spin of the Mu electron does not exhibit rapid fluctuations through spin exchange with the itinerant electrons, but is kept ``fixed" by the local ferromagnetic ordering mediated by itself (the electron and the four Mn neighbors are thought to be coupled and to behave as a single entity with large spin: i.e. a ``spin polaron"). 
In this scenario, the two frequencies correspond to two muon spin-flip transitions between states with fixed electron spin orientation.
The picture of such a localized electron state appears controversial given the metallic nature of MnSi.

The aim of the present paper is to lift the controversy regarding the occurrence or not of magnetic polaron and provide information on the muon stopping site. We retain this an essential point to solve, as this will show whether $\mu$SR can be utilized to specifically extract quantitative information on this heavily studied strongly correlated system. In this study, we will show that the occurrence of different frequencies in the 
TF-$\mu$SR spectra can be traced back to the different responses of crystallographically equivalent muon-stopping sites which become magnetically inequivalent in the presence of an external magnetic field and that one does not need to invoke a magnetic polaron to explain the TF-$\mu$SR data. We will also show that the occurrence of two spontaneous muon frequencies in the zero-field $\mu$SR data recorded below $T_C$ can be explained within this frame. Finally, we also present density-functional theory calculations which perfectly support the experimental evidences. 

\section{Experimental Details}
\begin{figure}[b]
\includegraphics[width=0.8\columnwidth]{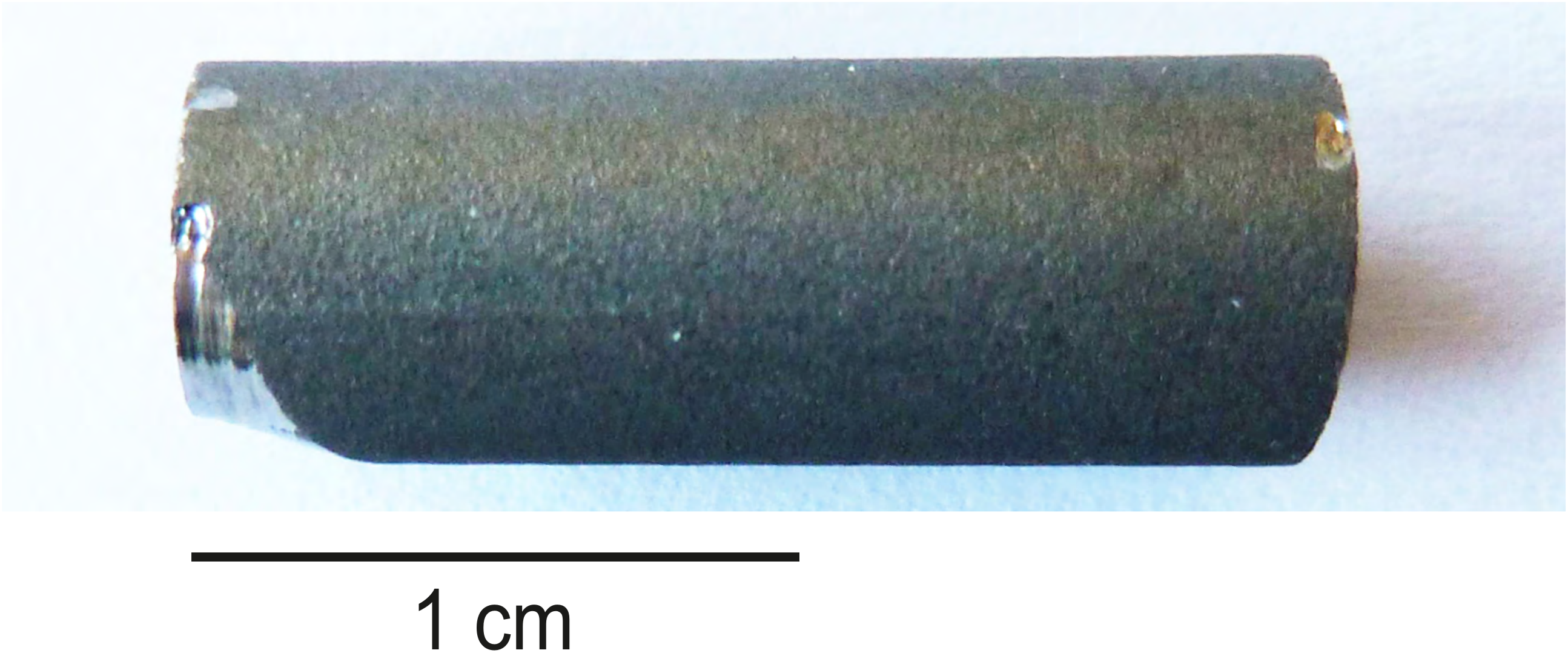}
\caption{Sample used for the TF-$\mu$SR measurements. The sample was rotated along the cylinder axis. 
} \label{Sample}
\end{figure}

We used the same MnSi sample as in the previous $\mu$SR under pressure experiments \cite{Andreica}. It consists of a single crystal with the form of a cylinder of  approximately 7 mm diameter and 19 mm length (see Fig.~\ref{Sample}).
The single crystal was grown by the Czochralsky pulling technique from a stoichiometric melt of high-purity elements ($>99.995$\%) using radio-frequency heating and a cold copper crucible. No deviation from the known crystal structure [space group $P2_{1}3$, No. 198; Mn-ion at the position $(0.138,0.138,0.138)$ and Si-ion at the position $(0.845,0.845,0.845)$; lattice constant 4.558~\AA] nor any
presence of foreign phase was detected by scanning electron microscope 
microanalysis, backscattered electron imaging, and x-ray Debye
Scherrer pattern.

The study of the angular dependence of the muon Knight-shift in MnSi was performed using the GPS instrument located at the $\pi$M3 beamline of the HIPA Complex at the Paul Scherrer Institut (PSI, Villigen, Switzerland). The measurements were performed with the transverse-field geometry with a magnetic field of 5200~G. The sample was rotated around the cylinder axis and the external field $\mathbf{B}_{\rm ext}$ was applied perpendicular to the cylinder. Note that the sample 
was not grown along a specified crystallographic orientation, i.e. the cylinder axis does not correspond to a principal axis. 
The measurements presented here were performed using a dynamical He-flow cryostat (i.e. sample in He-flow) at a temperature of 50~K. The temperature stability of the sample was better than 0.1~K for all the measurements.

\section{Results and Discussion}
\subsection{Transverse-Field Data and Muon Site Determination}
For all rotation angles $\phi'$, the $\mu$SR signals are best modeled by the presence of four frequencies with different angular dependences. Fits with all parameters free point to similar amplitudes for the four components. Hence, in a second series of fits the amplitudes were forced to be the same. That is:
\begin{equation}
\mathcal{A}_{\rm tot}G(t)=\sum_{i=1}^4\mathcal{A}\exp(-\lambda_i t)\cos(\nu_i\,2\pi\,t + \psi)~.
\label{asymtheor}
\end{equation}

\begin{figure}[b]
\includegraphics[width=0.9\columnwidth]{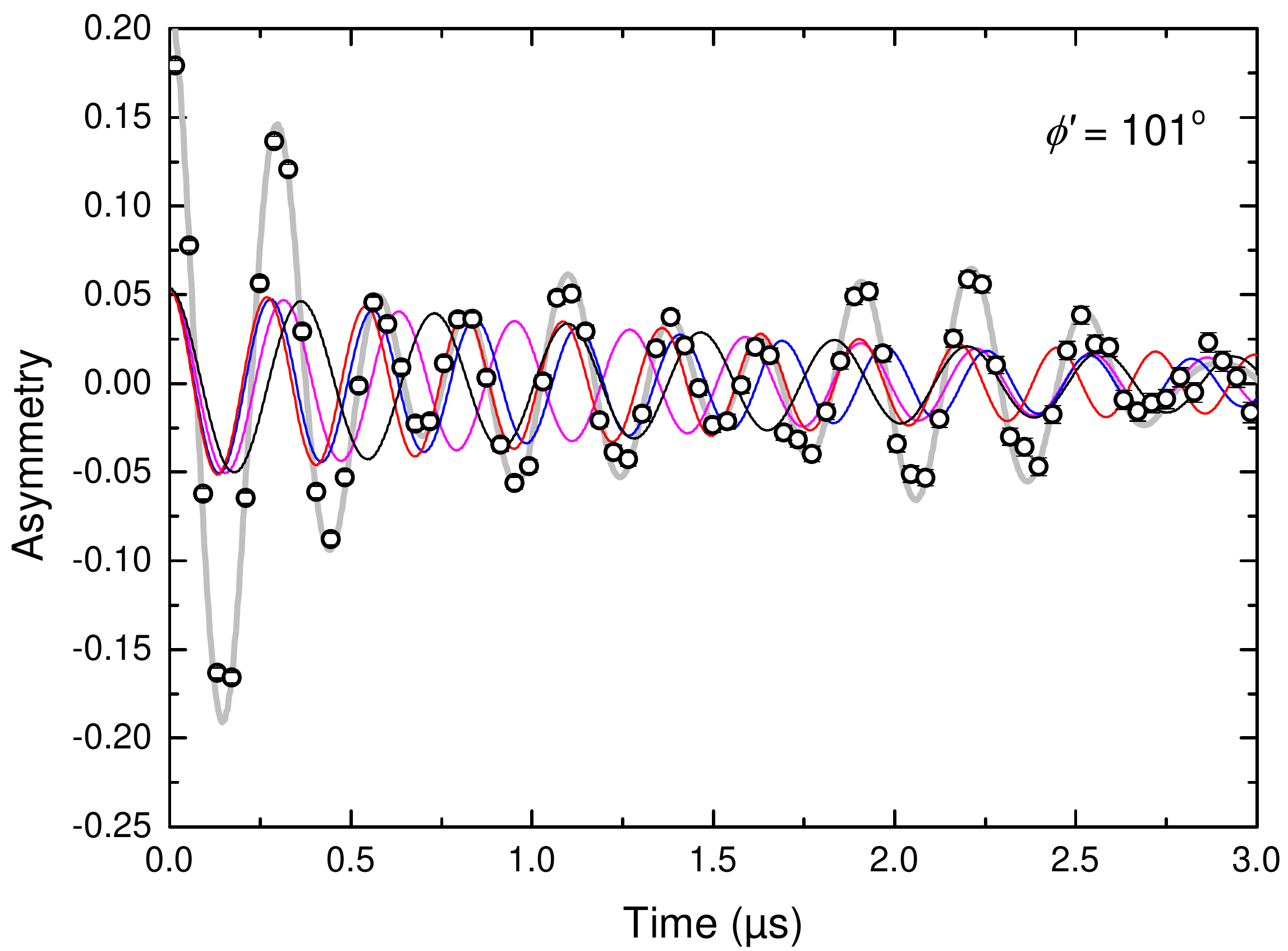}
\caption{(Color online) Example of one $\mu$SR time spectrum plotted in a rotating reference frame (RRF) with frequency 65 MHz. The grey line represents the fit to the data (open points) using Eq.~(\ref{asymtheor}) and the four individual signals are also displayed (color code is the same as on Fig.~\ref{Raw_data}).  Note that the actual fits were performed without RRF and without time binning (i.e. original bin of 0.9766~ns). Notice that the four signals have very similar depolarization rates.} \label{Raw_data_Time}
\end{figure}
\begin{figure}[t]
\includegraphics[width=0.75\columnwidth]{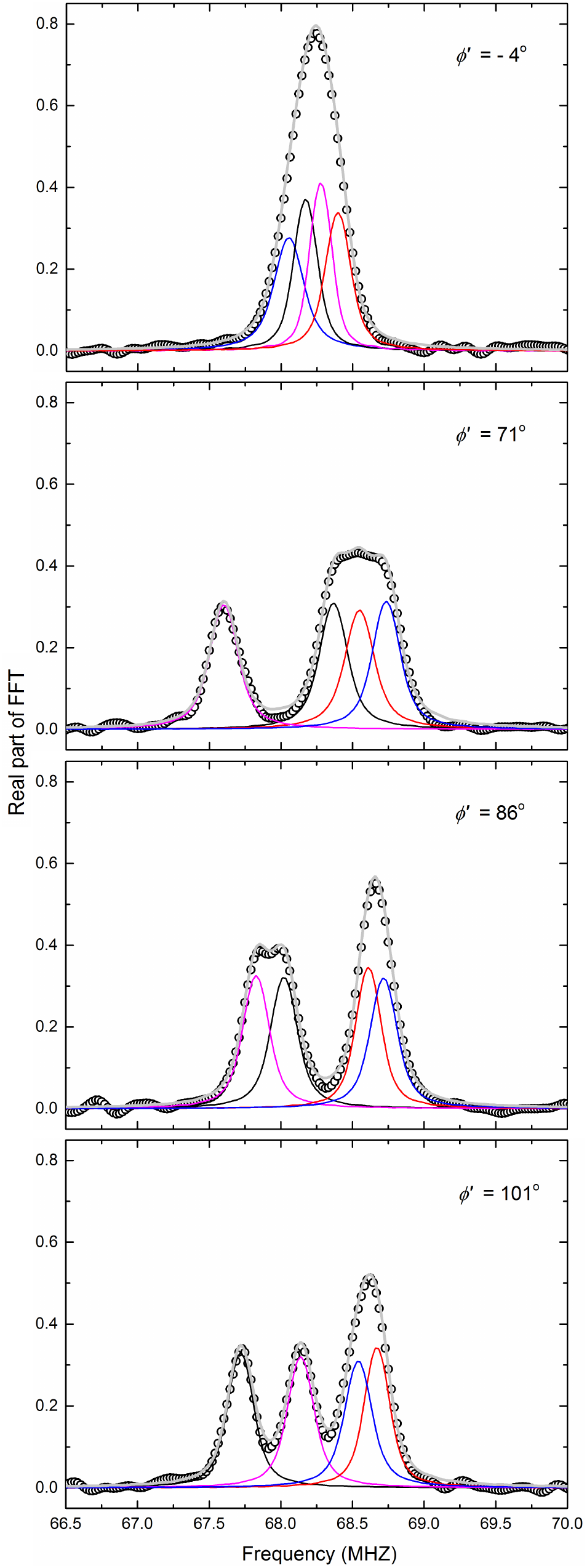}
\caption{(Color online) Fourier transform of TF spectra taken at different orientations. The color code of the different signals is the same as on Fig.~\ref{Raw_data_Time} and \ref{Raw_data}.} \label{Raw_data_FFT}
\end{figure}
\begin{figure}[t]
\includegraphics[width=0.9\columnwidth]{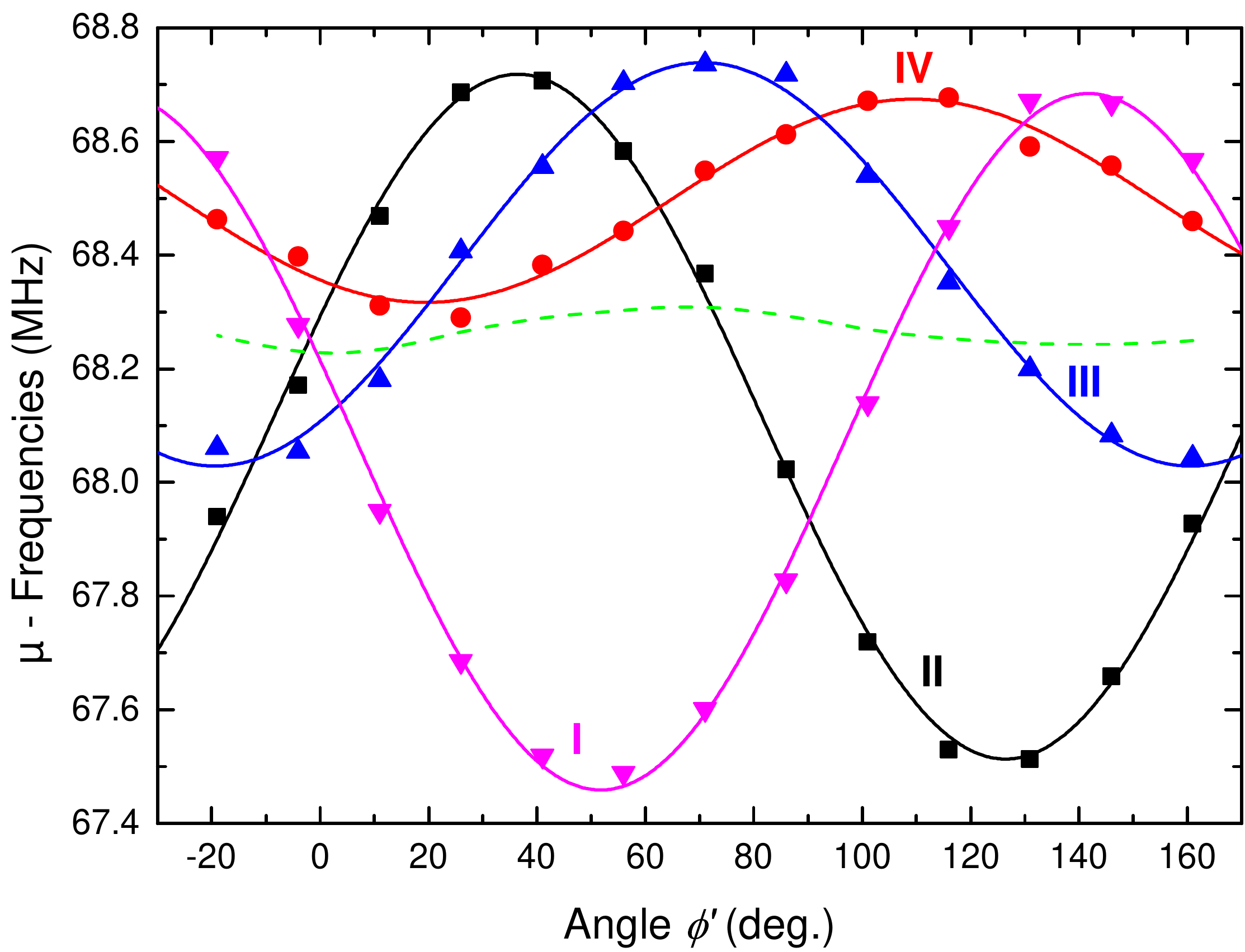}
\caption{(Color online) Angular dependence of the fitted $\mu$SR frequencies. The lines are guides to the eye. The green dash-line represents the average frequency $\bar{\nu}(\phi')$ (see text).} \label{Raw_data}
\end{figure}

The frequencies $\nu_i$ reflect the local magnetic field values sensed by the muon at the stopping site(s), i.e. $\nu_i = \gamma_{\mu}|{\mathbf B}_{\text{loc,\tiny TF},i}|/({2\pi})$, where $\gamma_{\mu}$ is the gyromagnetic ratio of the muon. The phase parameter $\psi$ is common to all signals as it is defined by the direction of the muon spin with respect to the detector system at the muon implantation time. 
During the fits, no restrictions were applied concerning the frequencies and the results were directly obtained through the software {\sc musrfit} \cite{Suter}. Note that all the fits were performed in the time-space between 0 and 8~$\mu$s. As illustrations to the data, we present in Fig.~\ref{Raw_data_Time} a $\mu$SR time-spectrum, plotted in a rotating reference frame, and in Fig.~\ref{Raw_data_FFT} selected Fast Fourier Transform (FFT) of spectra recorded at different angles. The full angular dependence of the frequencies is shown in Fig.~\ref{Raw_data}.

The occurrence of four frequencies in the $\mu$SR signal clearly points to a muon site located at the $4a$ Wyckoff position of the cubic structure. In this view, the main contribution of the observed angular dependence arises from the dipolar contribution. As shown in 
Table~\ref{diptensor}, which provides information on this type of sites and the respective dipolar tensor forms, taking into account the symmetry of the dipolar tensors, the four crystallographically equivalent 4$a$ sites in the unit cell are either: i) magnetically equivalent when $\mathbf{B}_{\rm ext}$ is applied along a principal direction; ii) grouped in two pairs of magnetically inequivalent sites for $\mathbf{B}_{\rm ext}$ applied in a principal plane; iii)~usually all magnetically inequivalent for $\mathbf{B}_{\rm ext}$ applied along an arbitrary direction.

\begingroup
\begin{table*}[t]
\caption{\label{diptensor}Coordinates and form of the representation of the dipolar tensor ${\mathbf A}_{\rm dip}$ in the crystal reference frame for the crystallographic Wyckoff position 4$a$ (four crystallographically equivalent sites designed here by $I$ to $IV$). We shall call $[A_{ij}^c]$ this representation. Note that the value $|a_{\rm dip}|$ is the same for all the tensors.\\
Bear also in mind that the structure can also accommodate Wyckoff positions of type 12$b$ (all other sites not equivalent to a 4$a$ position) but in this case one would expect to observe twelve frequencies in the $\mu$SR signal for an arbitrary direction of the applied magnetic field.}
\begin{ruledtabular}
\begin{tabular}{l|cccc}
 & site 4$a$-$I$ & site 4$a$-$II$ & site 4$a$-$III$ & site 4$a$-$IV$ \\
\hline~\\
Coordinates~~~& $(x,x,x)$&$(\tfrac{1}{2}-x,\bar{x},\tfrac{1}{2}+x)$&
$(\tfrac{1}{2}+x,\tfrac{1}{2}-x,\bar{x})$&$(\bar{x},\tfrac{1}{2}+x,\tfrac{1}{2}-x)$\\~\\
\hline~\\
$[A_{ij}^c]$~~~& $\begin{pmatrix} 0&a_{\rm dip}&a_{\rm dip}\\a_{\rm dip}&0&a_{\rm dip}\\a_{\rm dip}&a_{\rm dip}&0\end{pmatrix}$&$\begin{pmatrix} 0&a_{\rm dip}&-a_{\rm dip}\\a_{\rm dip}&0&-a_{\rm dip}\\-a_{\rm dip}&-a_{\rm dip}&0\end{pmatrix}$
&$\begin{pmatrix} 0&-a_{\rm dip}&-a_{\rm dip}\\-a_{\rm dip}&0&a_{\rm dip}\\-a_{\rm dip}&a_{\rm dip}&0\end{pmatrix}$
&$\begin{pmatrix} 0&-a_{\rm dip}&a_{\rm dip}\\-a_{\rm dip}&0&-a_{\rm dip}\\a_{\rm dip}&-a_{\rm dip}&0\end{pmatrix}$\\~\\
\end{tabular}
\end{ruledtabular}
\end{table*}
\endgroup

\begin{figure}[b]
\includegraphics[width=0.9\columnwidth]{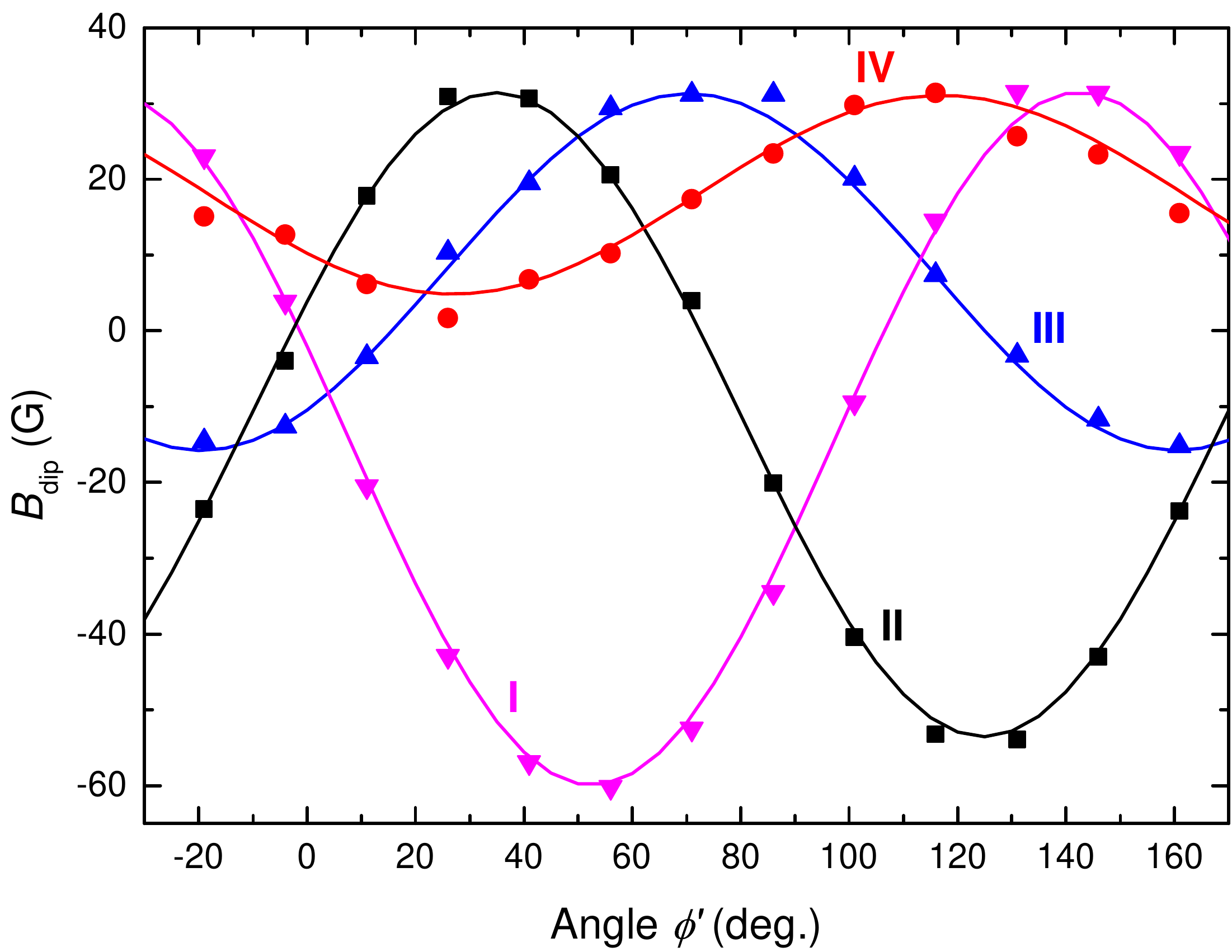}
\caption{(Color online) Angular dependence of the dipolar contributions of the four frequencies reported on Fig.~\ref{Raw_data}. The symbols are obtained after the subtraction described in Eq.~(\ref{demcorr}). The lines correspond to the theoretical calculations positioning the muon at the site $(0.532,0.532,0.532)$ and taking into account the rotation axis (see text).} \label{Bdip_corrected}
\end{figure}

Generally, the local field at the muon-site can be written as usual as (see for example Refs.~\onlinecite{YaouancDalmas, Amato})
\begin{equation}
{\mathbf B}_{\text{loc,\tiny TF}}(\phi') = {\mathbf B}_{\rm ext}+{\mathbf B}_{\rm cont} + {\mathbf B}_{\rm dip}(\phi')+{\mathbf B}_{\rm Lor}+{\mathbf B}_{\rm dem}
(\phi')~,
\label{Eq_Bloc}
\end{equation}
where the values of ${\mathbf B}_{\rm ext}$, of the contact term ${\mathbf B}_{\rm cont}$ and of the Lorentz field ${\mathbf B}_{\rm Lor}$ are independent of the rotation angle $ \phi'$  (bear in mind that for a cubic system, the representation of the magnetic susceptibility tensor in the crystal reference frame is given by $\bm{\chi}=\chi\,{\mathbf E}$, where ${\mathbf E =[E_{ij}]=[\delta_{ij}]}$). We mention at this point that, as said, we expect that the essential part of the observed angular dependence should arise from the dipolar contribution. Nevertheless, we cannot exclude a small angular dependence of the demagnetization field, as the sample had not a perfect cylindrical shape (see Fig.~\ref{Sample}). A way out is to observe that i) for a given field direction the demagnetization field is identical for all signals; ii) due to the symmetry of the dipolar tensors for the 4$a$ sites, the sum of the dipolar contribution of the four signals should always cancel (see Table~\ref{diptensor}). Hence the true angular dependence of the measured dipolar contribution for each signal can be obtained by
\begin{equation}
B_{{\rm dip}, i} (\phi') = \frac{(\nu_i(\phi') - \bar{\nu}(\phi'))\,2\pi}{\gamma_{\mu}}~,
\label{demcorr}
\end{equation}
where $\bar{\nu}(\phi')$ is the average frequency at each angle. We observe that the amplitude of the angular dependence of  $\bar{\nu}(\phi')$ is very small, i.e. it represents a field $\bar{\nu}(\phi')2\pi/\gamma_{\mu}$ of the order of 1.8~G, which would correspond to a variation of less than 3\% of the demagnetization factor during the rotation (see Fig.~\ref{Raw_data}). The average frequency $\bar{\nu}(\phi')$ can be interpreted as the sum of all other contributions, that is
\begin{equation}
\bar{\nu}(\phi') = \frac{\gamma_{\mu}}{2\pi}\,|{\mathbf B}_{\rm ext}+{\mathbf B}_{\rm cont} + {\mathbf B}_{\rm Lor}+{\mathbf B}_{\rm dem}(\phi')|
\end{equation}
For the four signals, the dipolar contributions obtained with the slight correction described in Eq.~(\ref{demcorr}) are shown on Fig.~\ref{Bdip_corrected}. 

\begin{figure}[t]
\includegraphics[width=1\columnwidth]{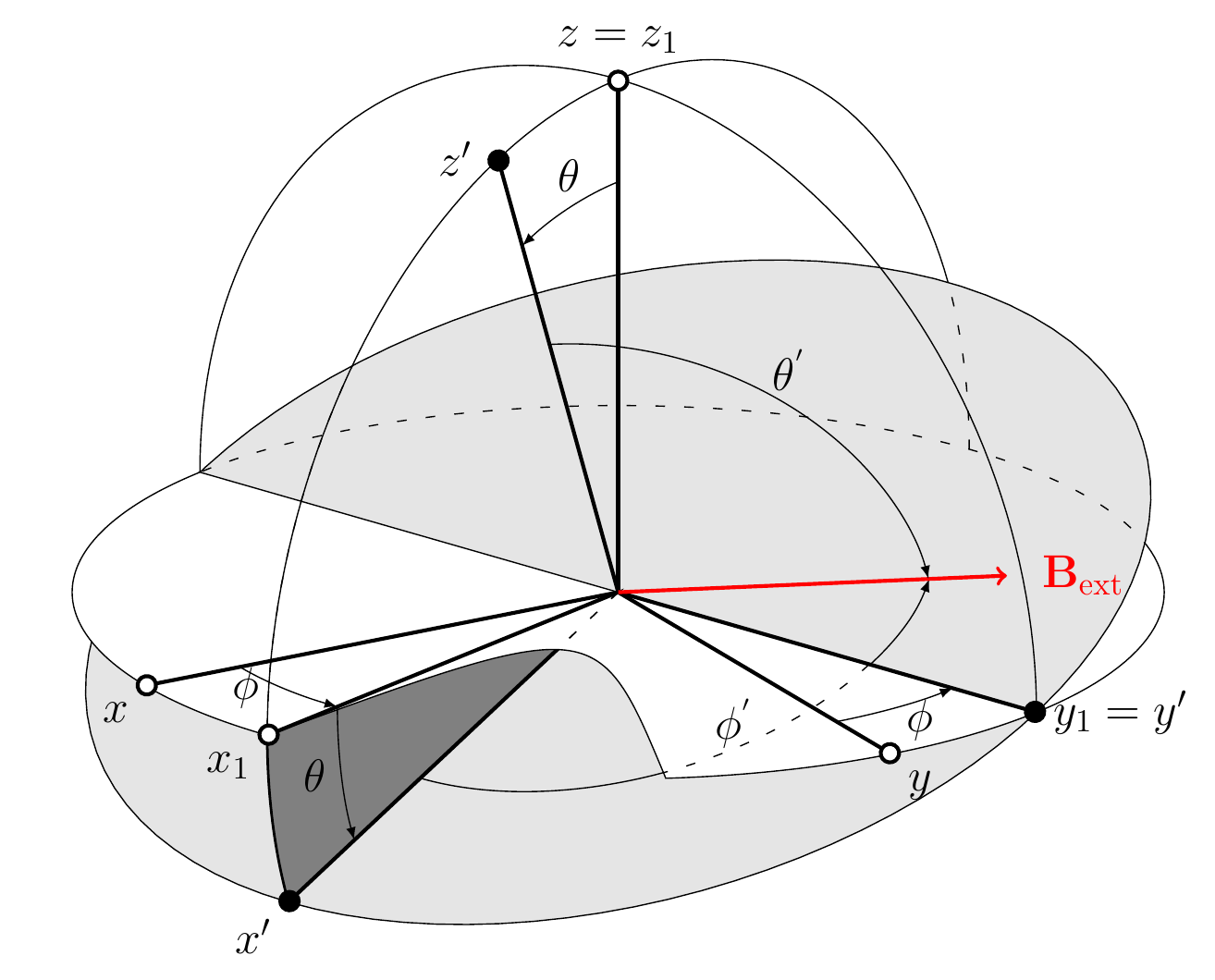}
\caption{(Color online) Definition of the Euler angles defining the rotation axis and of the polar and azimuth angles defining the direction of the external field. The reference frame $(x,y,z)$ is the reference frame of the crystal and $(x',y',z')$ is the one defined by the rotating plane and rotation axis.} \label{Euler_Angles}
\end{figure}

In the following, we will carefully analyze the angular dependence of these dipolar contributions to determine the muon-stopping site. The dipolar field contribution to the value of the local field sensed by the muon can be written as
\begin{equation}
{\mathbf B}_{\rm dip}= {\mathbf A}_{\rm dip}\,\bm{\chi}\,{\mathbf B}_{\rm ext}~,
\end{equation}
where ${\mathbf A}_{\rm dip}$ is the dipolar tensor 
and $\bm{\chi}\,{\mathbf B}_{\rm ext}$ represents the local Mn-moment induced by the external field.
As in the present experiment the field was not rotated in a principal crystallographic plan, we will define $[A_{i'j'}^l]$ as the representation of the dipolar tensor in the reference frame given by the rotation axis and rotation plane of the applied field (i.e. the laboratory frame, see Fig.~\ref{Euler_Angles}). Expressing the rotation axis with the Euler angles $\theta$ and $\phi$, we have the relation
\begin{equation}
[A_{i'j'}^l]= R_{[010],\theta}\,R_{[001],\phi}\,[A_{ij}^c]\,R_{[001],-\phi}\,R_{[010],-\theta}~,
\label{eqrot}
\end{equation}
where the rotation matrices $R_{\bm{\alpha},\beta}$ represent a rotation of angle $\beta$ around the direction $\bm{\alpha}$ and $[A_{ij}^c]$ is the representation of the dipolar tensor in the crystal reference frame (see Table~\ref{diptensor}). In the reference frame of the rotation axis and plane, we can express the external field as ${\mathbf B}_{\rm ext}=B_{\rm ext}(\sin \theta'\cos \phi',\sin \theta'\sin \phi',\cos \theta')$, where $\theta'$ and $\phi'$ are now the polar and azimuth angles of the field in the new reference frame (see Fig.~\ref{Euler_Angles}). Generally, the angular dependence of the component of ${\mathbf B}_{\rm dip}$ along the direction of the external field can be expressed as \cite{Feyerherm}
\begin{eqnarray}
B_{\rm dip,||}\! & =\! & \tfrac{1}{3}(A_{x'x'}^l\chi_{x'}^l\!+\!A_{y'y'}^l\chi_{y'}^l\!+\!A_{z'z'}^l\chi_{z'}^l) B_{\rm ext}\nonumber \\
\,&\,&+\tfrac{2}{3}[A_{z'z'}^l\chi_{z'}^l\!-\!\tfrac{1}{2}(A_{x'x'}^l\chi_{x'}^l\!+\!A_{y'y'}^l\chi_{y'}^l)]\,P_2^0(\cos \theta') B_{\rm ext}\nonumber \\
\,&\,&-\tfrac{1}{3}A_{x'z'}^l(\chi_{x'}^l\!+\!\chi_{z'}^l)\,P_2^1(\cos \theta')\cos \phi'B_{\rm ext}\nonumber \\
\,&\,&-\tfrac{1}{3}A_{y'z'}^l(\chi_{y'}^l\!+\!\chi_{z'}^l)\,P_2^1(\cos \theta')\sin \phi'B_{\rm ext}\nonumber \\
\,&\,&+\tfrac{1}{6}(A_{x'x'}^l\chi_{x'}^l\!-\!A_{y'y'}^l\chi_{y'}^l)\,P_2^2(\cos \theta')\cos 2\phi'B_{\rm ext}\nonumber \\
\,&\,&+\tfrac{1}{6}A_{x'y'}^l(\chi_{x'}^l\!+\!\chi_{y'}^l)\,P_2^2(\cos \theta')\sin 2\phi'B_{\rm ext}~, \label{Adipgeneral}
\end{eqnarray}
where 
the terms $P_l^m$ represent the usual associated Legendre polynomials \cite{Amato_foot_1}.

Equation~(\ref{Adipgeneral}) can be simplified by noting that for a cubic system 
the magnetic susceptibility is isotropic and here defined as $\chi$.
We also have $\theta'=90^{\circ}$, as of course our field is per definition perpendicular to the rotation axis and $\phi'$ represents our rotation angle. We therefore have
\begin{eqnarray}
B_{\rm dip,||} & = & \tfrac{1}{2}\,\chi\,(A_{x'x'}^l+A_{y'y'}^l) B_{\rm ext}\nonumber \\
\,&\,&+\,\tfrac{1}{2}\,\chi\,(A_{x'x'}^l-A_{y'y'}^l)\,\cos 2\phi'\, B_{\rm ext}\nonumber \\
\,&\,&+\,\chi\,A_{x'y'}^l\,\sin 2\phi'\, B_{\rm ext}~.\label{Adipparticul}
\end{eqnarray}
Note that the local field at the muon site ${\mathbf B}_{\text{loc,\tiny TF}}$ is not exactly parallel to ${\mathbf B}_{\rm ext}$ as in general the dipolar tensor is not diagonal. However as $|{\mathbf B}_{\rm ext}+{\mathbf B}_{\rm cont} +{\mathbf B}_{\rm Lor}+{\mathbf B}_{\rm dem}| \gg |{\mathbf B}_{\rm dip}|$, it is enough to consider solely the component of ${\mathbf B}_{\rm dip}$ along the direction of the external field [the differences are here of the order of 0.1\% on the values of $B_{\rm dip}$ obtained by Eq.~(\ref{demcorr})]. 

As $B_{\rm ext}$ (5200~G) and $\chi$ (0.030~emu/mole \cite{suscep1,suscep2,suscep3}) are known, the task is now to obtain the elements $A_{x'x'}^l$, $A_{y'y'}^l$ and $A_{x'y'}^l$ by fitting the angular dependence of each muon frequency \cite{Amato_foot_2}.  
The results of the fits obtained by adjusting Eq.~(\ref{Adipparticul}) on the angular dependence of each signal are shown on Table~\ref{table_fits}.

\begin{table}[t]
\caption{\label{table_fits}Fit results obtained adjusting Eq.~(\ref{Adipparticul}) on the angular dependence of the dipolar contribution for each signal.}
\begin{ruledtabular}
\begin{tabular}{l|ccc}
 & $A_{x'x'}^l$   & $A_{y'y'}^l$ & $A_{x'y'}^l$ \\
 & (mole/emu) & (mole/emu) & (mole/emu) \\
\hline~\\
Signal $I$~~~& -0.0128(38)&-0.1713(11)&-0.2925(31)\\
Signal $II$~~~& 0.0235(30)&-0.1662(32)&0.2617(26)\\
Signal $III$~~~& -0.0640(38)&0.1747(35)&0.0945(27)\\
Signal $IV$~~~& 0.0538(48)&0.1627(51)&-0.0637(41)\\~\\
\end{tabular}
\end{ruledtabular}
\end{table}

\begin{figure}[t]
\includegraphics[width=0.9\columnwidth]{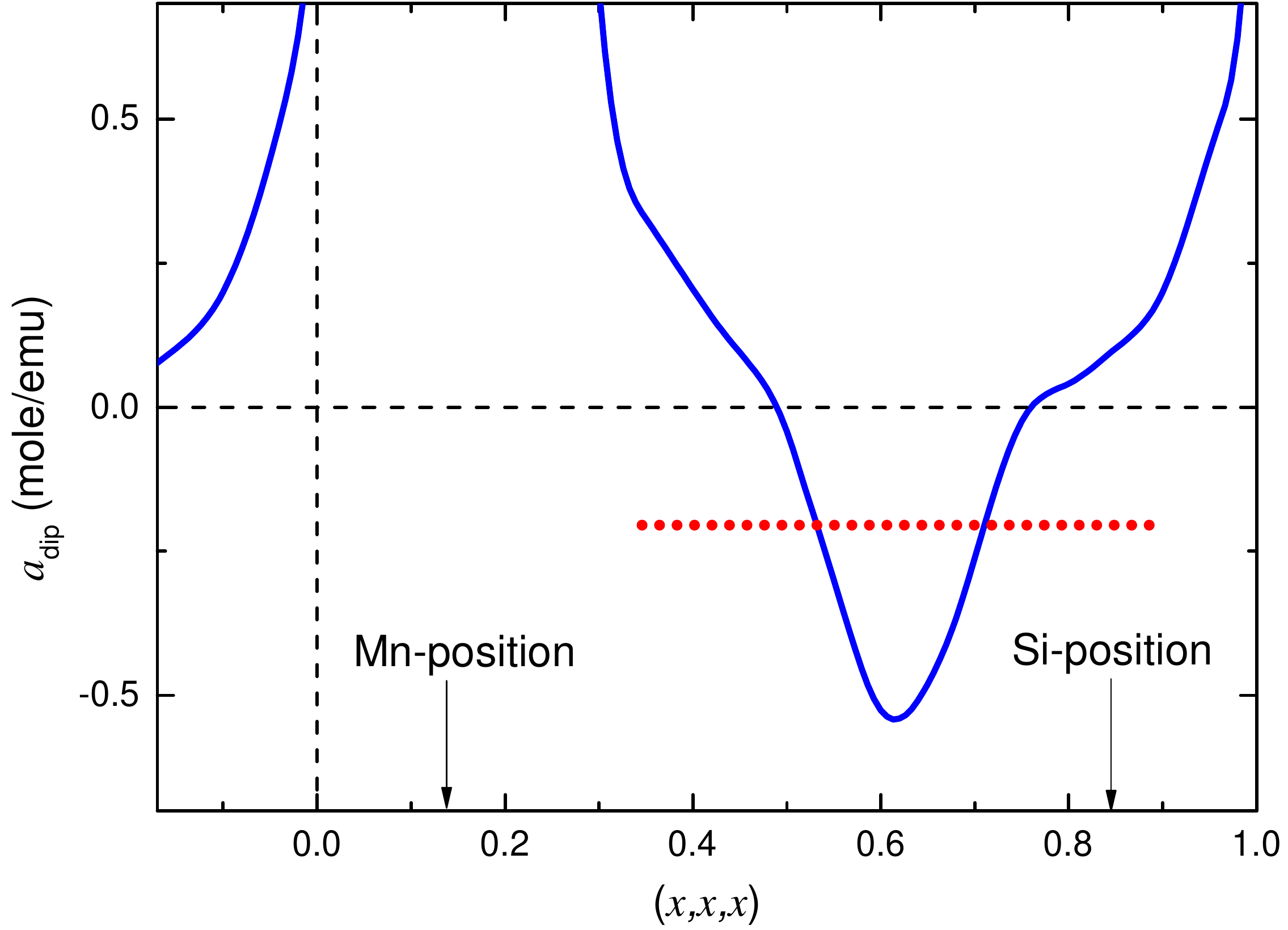}
\caption{(Color online) Dipolar sum calculation of the parameter $a_{\rm dip}$ characterizing the representation $[A_{ij}^c]$ of the dipolar tensors in the crystal reference frame for the 4$a$ sites. The divergence occurs at the 4$a$ position of the Mn-ion. The red dot-line corresponds to the results of the fits as explained in the text.} \label{Adip_theor}
\end{figure}

\begin{figure}[t]
\includegraphics[width=0.85\columnwidth]{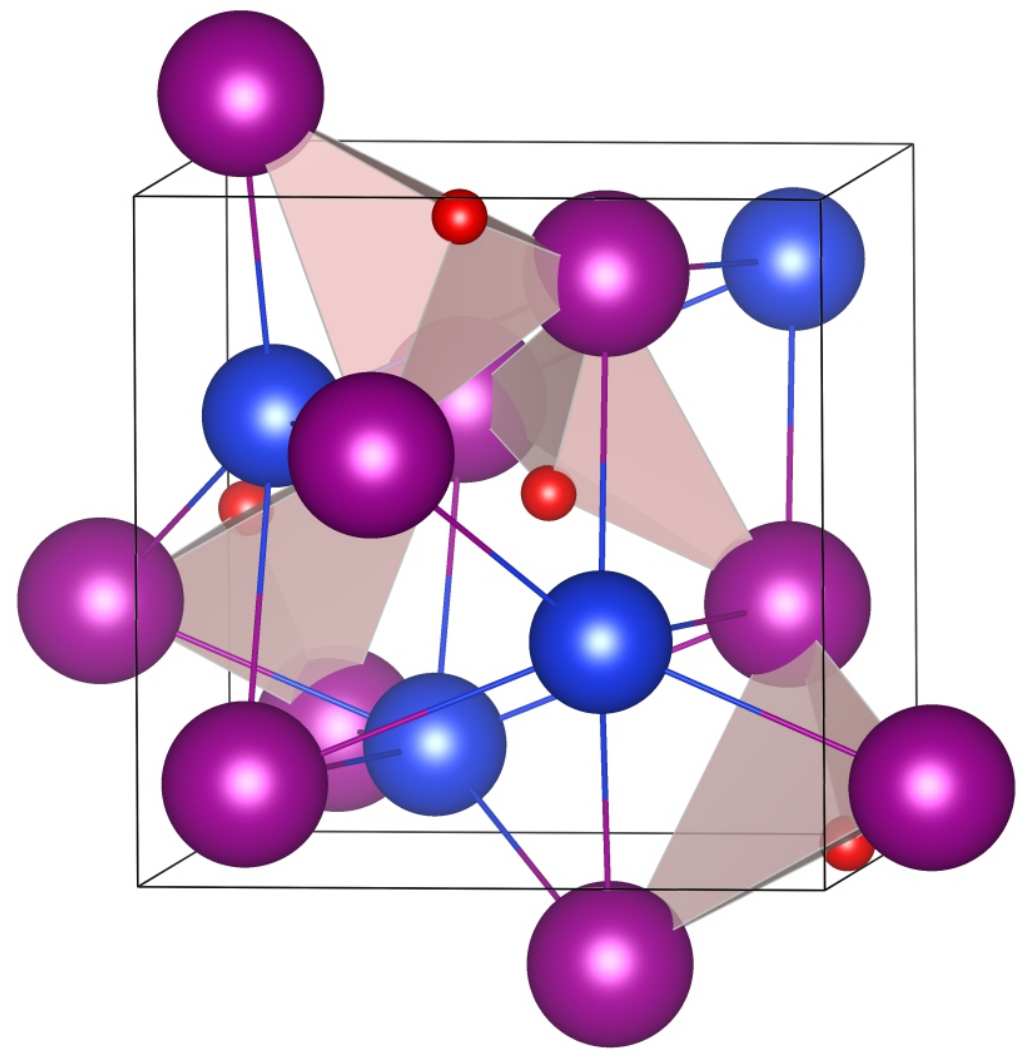}
\caption{(Color online) Sketch of the crystallographic structure of MnSi (Mn-ions are drawn in purple, Si-ions in blue). The muon position $(0.532,0.532,0.532)$ is also indicated (red) as well as the other three equivalent sites. Note that six Mn-ions , which do not belong to the primary unit cell, are also displayed.} \label{Structure}
\end{figure}

At this point, these results can be used to solve Eq.~(\ref{eqrot}) for each signal (site) and obtain the Euler angles $\theta$ and $\phi$ of the rotation axis, as well as the parameter $a_{\rm dip}$ characterizing the representation of the dipolar tensors ${\mathbf A}_{\rm dip}$ in the reference frame of the crystal (see Table~\ref{diptensor}). As each signal should provide us with the same parameters, we have a rather precise determination of these parameters, i.e. $a_{\rm dip}=-0.2044(40)$~mole/emu, $\theta=83(1)^\circ$ and $\phi=242(1)^\circ$. We note that the values of $\theta$ and $\phi$ could be confirmed {\it a posteriori} by performing x-rays Laue measurements of the crystal. But we stress that the values of $\theta$ and $\phi$ extracted by $\mu$SR must be considered as more precise as, for example, they take into account any possible misalignment of the sample mounting compared to the direction of the external magnetic field. The experimentally determined value of $a_{\rm dip}$ can be directly compared with dipolar sum calculations. Figure~\ref{Adip_theor}, which exhibits the theoretical value of $a_{\rm dip}$ calculated for different sites 4$a$, indicates that this experimental value is compatible with muons sitting either at the site $(0.532,0.532,0.532)$ or  $(0.721,0.721,0.721)$. 
We stress that a small uncertainty on the magnetic susceptibility (and therefore finally on the obtained value of $a_{\rm dip}$) will only weakly affect the determination of the muon-site (for example, an uncertainty as high as $\pm$10\% on $\chi$ leads to a shift of $\Delta x=\pm0.005$ on the muon-site coordinates).  
The obtained sites correspond to almost symmetrical positions along the crystal diagonal on each side of the plane formed by three adjacent Mn-ions (see Fig.~\ref{Structure}). However, the site $(0.721,0.721,0.721)$ appears not probable as located rather close  to the Si-ion on the crystal diagonal ($r_{\mu - {\rm Si}} \simeq 1.0$~\AA). Note that such a short $\mu$-ion distance can sometimes be observed when a bounding between the muon and a negative ion occurs, as in the high-$T_c$ cuprates. 

To demonstrate the perfect agreement between theoretical calculations and experiment, the lines on Fig.~\ref{Bdip_corrected} represent the dipolar calculations for the site $(0.532,0.532,0.532)$ with the Euler angles $\theta=83^\circ$ and $\phi=242^\circ$. Note that as the four sites are crystallographically equivalent, one expects similar amplitudes of the four signals, as experimentally observed.   We note also that our measurements performed with angles of $\phi'=86^\circ$ and $-4^\circ$ correspond respectively to an external field applied almost into a principal plane [i.e. (110)] and almost along the principal axis $[00\bar{1}]$ (as the Euler angle for the rotation axis is $\theta=83^\circ$). Figure~\ref{Raw_data_FFT} shows that the signals tend to collapse  into two pairs in the first case and all together (and with a dipolar contribution tending to zero) in the second case, as expected.   For completeness, we would also like to mention the possibility of a small lattice dilation around the muon. We notice first that such effects are usually extremely small in metals (of the order of the percent, if detectable at all; see for example Ref.~\onlinecite{Luke}). A hypothetical weak dilation of the lattice would decrease the calculated absolute values of the $a_{\rm dip}$ parameter, which would push slightly the muon position in the direction of the plane formed by the three adjacent Mn-ions (see Fig.~\ref{Structure}). However, the main conclusion that the muon is stopped at a 4$a$ position would not be affected.

Finally, we note that a by-product of our analysis is the determination of the contact field ${\mathbf B}_{\rm cont}$. Basically two interactions contribute to the contact field. The first one is a result of the Pauli paramagnetism of the conduction electrons and their Fermi contact interaction with the muon \cite{YaouancDalmas}. In systems with localized moments, as MnSi, an additional contribution arises as the spin-polarization of the conduction electrons at the muon site will be increased through the Ruderman-Kittel-Kasuya-Yosida interaction due to the local magnetic moments. As the Pauli susceptibility of the conduction electrons is much smaller than the magnetic susceptibility from the local $3d$ moments (i.e. $\chi \simeq \chi_{3d} \gg \chi_{\rm Pauli}$), the Fermi interaction can be safely neglected in the contact field, which therefore can be written as 
\begin{equation}
{\mathbf B}_{\rm cont} = {\mathbf A}_{\text{cont,\tiny TF}}\,\bm{\chi}\,{\mathbf B}_{\rm ext}~,
\end{equation}
where ${\mathbf A}_{\rm cont}$ is the hyperfine contact coupling tensor. By writing Eq.~(\ref{Eq_Bloc}), we note that we implicitely assumed that the contact field is independent of the direction of ${\mathbf B}_{\rm ext}$ (i.e. ${\mathbf A}_{\text{cont,\tiny TF}} = A_{\text{cont,\tiny TF}}{\mathbf E}$) as observed in a number of compounds with local moments (see for example Ref.~\onlinecite{Amato_RMP} and references therein). In any case, even if one assumes that the weak angular dependence of $\bar{\nu}$ is solely arising from a hypothetical angular dependence of ${\mathbf A}_{\rm cont}$, this would correspond to a variation of less than 4\% of the hyperfine contact coupling tensor during the rotation. As we {\it know} that some angular dependence has to arise from the demagnetization factor (see Fig.~\ref{Sample}), this number has to be considered as an upper limit. By assuming that the hyperfine contact coupling tensor is isotropic, and taking into account that for our sample geometry the average value of the demagnetization factor is $\bar{N}=0.47\cdot(4\pi)$ one obtains a value of $A_{\text{cont,\tiny TF}} = -0.9276(20)$~mole/emu, which can now be included in the computation of the expected values of the spontaneous fields occurring at the muon sites below $T_C$ as described in the next Section. 

\subsection{Zero-Field Data Discussion}
At this point, having a solid knowledge of the muon stopping site, it appears legitimate to thoroughly discuss the number and values of the spontaneous $\mu$-frequencies observed below $T_C$ \cite{Kadono,Uemura,Andreica} in zero-applied field (ZF) $\mu$SR experiments. All the $\mu$SR studies reported to date reveal the occurrence of two spontaneous $\mu$-frequencies with values $\nu_{\text{\tiny ZF},1} \simeq 12.3$~MHz and $\nu_{\text{\tiny ZF},2} \simeq 28.0$~MHz for $T\rightarrow 0$~K. Our first task here will be to discuss why solely two frequencies are observed.  

The magnetic structure of MnSi is characterized by spins forming a left-handed incommensurate helix with a propagation vector $k\simeq 0.036$~\AA $^{-1}$ in the [111] direction \cite{Ishikawa_1,Ishikawa_2,Ishida}. The static Mn-moments ($\sim\!0.4~\mu_{\rm B}$ for $T \rightarrow 0$~K) point in a plane perpendicular to the propagation vector. The period, which is incommensurate to the lattice constant, is about 18~nm. Due to the incommensurability of the magnetic structure, one expects a continuous set of local fields at our 4$a$ sites, and therefore the field distribution $D_i(B_{\text{loc,\tiny ZF}})$ at each site must be considered (here the index $i$ distinguishes the four sub-sites 4$a$). It was shown that such a magnetic structure leads to a field distribution given by \cite{Andreica_Thesis,YaouancDalmas,Schenck}  
\begin{equation}
D_i(B_{\text{loc,\tiny ZF}})=\frac{2}{\pi}\frac{B_{\text{loc,\tiny ZF}}}{\sqrt{B_{\text{loc,\tiny ZF}}^2-B_{{\rm min},i}^2}\,\,\sqrt{B_{{\rm max},i}^2-B_{\text{loc,\tiny ZF}}^2}}~,
\label{eq_fd}
\end{equation}
and is characterized by two peaks due to the minimum and maximum cutoff field values. Hence, at a first glance, one would expect to observe up to eight peaks in the Fourier spectra of the ZF-$\mu$SR data. However, as the sites $II$, $III$ and $IV$ are located symmetrically around the direction [111] of the propagation vector (which represents for them a 3-fold symmetry axis), it can be shown that muons stopping at these sites will sense identical field distributions, given by Eq.~(\ref{eq_fd}), i.e. $D_{\text{\tiny\it II}}(B_{\text{loc,\tiny ZF}})=D_{\text{\tiny\it III}}(B_{\text{loc,\tiny ZF}})=D_{\text{\tiny\it IV}}(B_{\text{loc,\tiny ZF}})$. On the other hand, muons stopping at the site $I$ will all sense a unique field value irrespective to the phase of the helix at the muon stopping site. Hence, purely geometrical considerations already reduce the maximum number of peaks in the Fourier spectra of the ZF-$\mu$SR data down to three. 
Note that this conclusion is not affected if the sample is not monodomain. Hence for the other possible magnetic domains (which, by taking into account the helicity, are characterized by propagation vectors along [$\bar{1}\bar{1}1$], [$\bar{1}1\bar{1}$] and [$1\bar{1}\bar{1}$]) one obtains the same field distributions as for the original domain, albeit the symmetry of the sites with respect to the magnetic structure will be accordingly permuted. 

\begin{figure}[b]
\includegraphics[width=0.8\columnwidth]{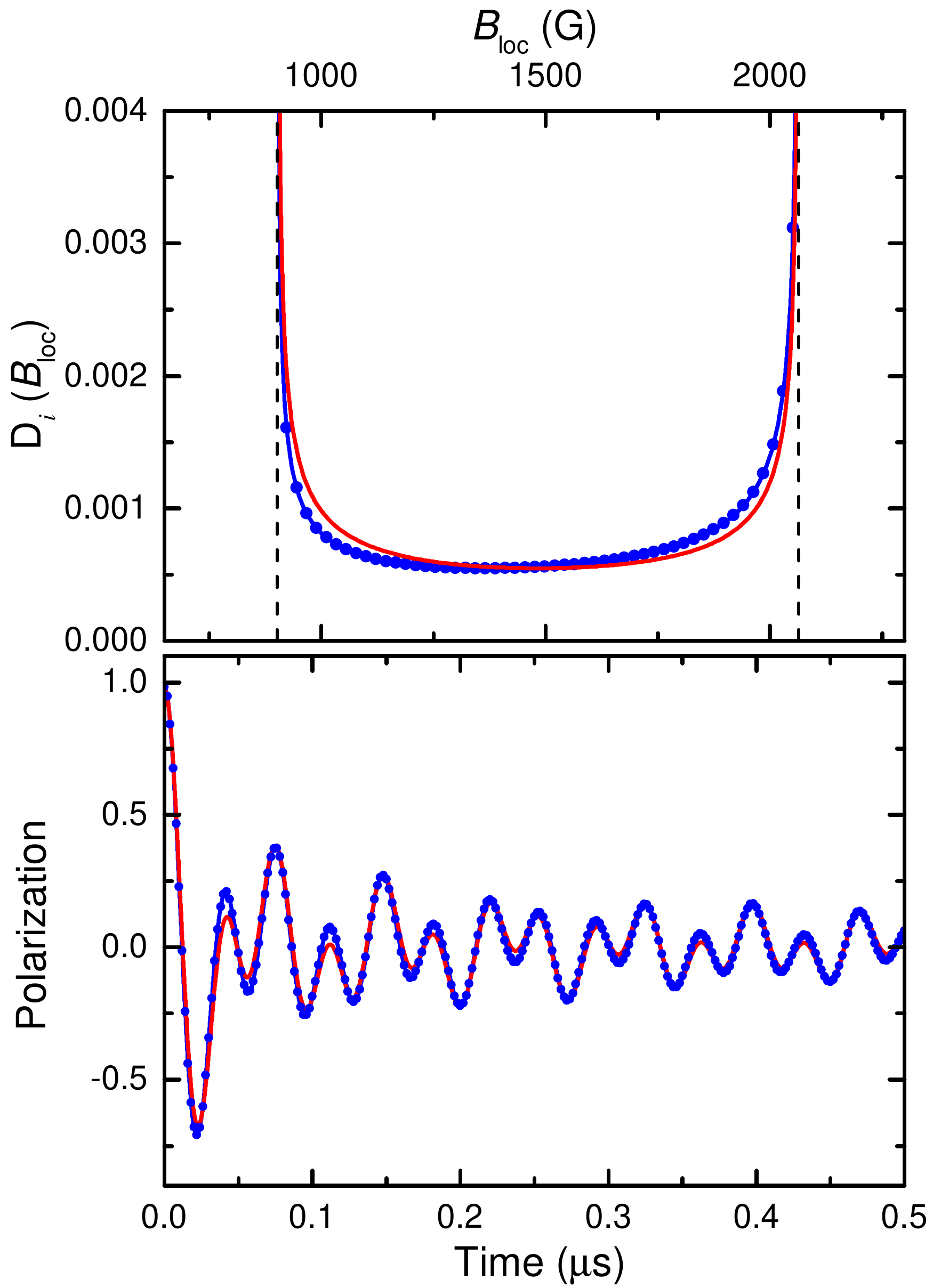}
\caption{(Color online) Upper panel: exact [blue line/symbols: see Eq.~(\ref{eq_fd})] and approximated [red line: see Eq.~(\ref{eq_fd_approx})] field distribution expected at the sites $II$, $III$ and $IV$. Lower panel: muon polarization function deduced from the field distributions shown on the upper panel (see also text).} \label{fig_approx_Bessel}
\end{figure}
To gain more insight, we have computed the expected theoretical cutoff field values of the field distributions. For a given type of muon site, the spontaneous local field in the incommensurate magnetic phase is given by\cite{Amato_foot_3}
\begin{equation}
\mathbf{B}_{\text{loc,\tiny ZF}}(\mathbf{r}^\prime_i) = \mathbf{B}_{\text{dip,\tiny ZF}}(\mathbf{r}^\prime_i)+\mathbf{B}_{\text{cont,\tiny ZF}}(\mathbf{r}^\prime_i)~.
\label{eq_B_loc_ZF}
\end{equation}
As before, the index $i$ distinguishes the four sub-sites 4$a$ and runs from $I$ to $IV$. 
Since the magnetic structure of MnSi is incommensurate, the spontaneous field differs from one particular site of the type $i$ to another site of the same type. In Eq.~(\ref{eq_B_loc_ZF}) $\mathbf{r}^\prime_i$ represents the set of vectors defining the positions of the sites of type $i$ with $\mathbf{r}^\prime_i=\mathbf{r}_i+\mathbf{R}$, where $\mathbf{r}_i$ defines the position of the site of type $i$ in the primitive cell and $\mathbf{R}$ is a vector belonging to the direct crystal lattice. The calculation of the dipolar contribution on the right-hand side of Eq.~(\ref{eq_B_loc_ZF}) is straightforward and will fully be determined by the knowledge of the muon stopping site and the details of the magnetic structure, such as propagation vector, value and direction of the magnetic moments, as well as helicity of the helix. On the other hand, the contact contribution can be written as
\begin{equation}
 \mathbf{B}_{\text{cont,\tiny ZF}}(\mathbf{r}^\prime_i)=V_{\text{mole}}A_{\text{cont,\tiny ZF}}\,\mathbf{M}(\mathbf{r}^\prime_i)~,
\label{eq_ZF_cont}
\end{equation}
where $V_{\text{mole}}$ is the volume of one mole of Mn-ions. In writing Eq.~(\ref{eq_ZF_cont}), we have assumed that the contact coupling is isotropic as suggested from our TF-data. 
$\mathbf M(\mathbf{r}^\prime_i)$ is the {\it local} magnetization at the muon site $\mathbf{r}^\prime_i$.
To compute $\mathbf{B}_{\text{cont,\tiny ZF}}$, the first natural choice is to set $A_{\text{cont,\tiny ZF}}$ equal to $A_{\text{cont,\tiny TF}}$ determined above. This is reasonable assuming that no massive changes occur on the Fermi surface when crossing $T_C$ and considering that due to the long wavelength, the close Mn-neighbors around the muon, which are determining the strength of the RKKY interaction, are essentially ferromagnetically aligned in the magnetic phase. 
This is the only choice made for the calculation, as the {\it local} magnetization $\mathbf{M}(\mathbf{r}^\prime_i)$ is also fully determined by the details of the magnetic structure and is given by
\begin{equation}
\mathbf M(\mathbf{r}^\prime_i)=\frac{1}{V_{\rm nn}}\sum_{j=1}^N{\mathbf m}_j =\frac{4}{NV_{\rm cell}}\sum_{j=1}^N{\mathbf m}_j~,
\label{eq_Mloc}
\end{equation}
where the volume $V_{\rm nn}=(NV_{\rm cell})/{4}$ is the volume occupied by the $N$ Mn-ions involved in the sum. The direction of the Mn-moments ${\mathbf m}_j$ is determined by the details of the magnetic structure. 
For our calculations, as the RKKY interaction is local, we have restricted the sum to a sphere having a radius of one lattice constant around the muon site.

The values of $\mathbf{B}_{\text{loc,\tiny ZF}}(\mathbf{r}^\prime_i)$ are located on the path of an ellipse for which the semimajor and semiminor axis values are $B_{\rm max}$ and $B_{\rm min}$. Starting from a known magnetic structure, the details for a precise computation of these values is thoroughly  explained in Ref.~\onlinecite{YaouancDalmas}.  The present calculations where performed on a sphere containing more than $5 \times 10^5$ unit cells. By setting $A_{\text{cont,\tiny ZF}}=A_{\text{cont,\tiny TF}}$, we stress again that our calculation does not contain any free parameters as the magnetic structure is taken from the literature, the muon site is determined by our TF-data and the contact coupling constant is taken {\it as is} from the TF-data. From this calculation, one obtains $B_{\text{max,\tiny\it II,III,IV}}= 2065$~G (corresponding to a muon frequency of 28.0~MHz), $B_{\text{min,\tiny\it II,III,IV}}= 900$~G (12.2~MHz) and $B_{\text{\tiny\it I}}= 880$~G (11.9~MHz).
The lower panel of Fig.~\ref{fig_simul_ZF} shows the obtained field distributions which have been convoluted for a better visibility (see caption). The red contribution represents the sum of the field distributions for the $II$, $III$ and $IV$ sites whereas the blue contribution shows the field distribution expected at the $I$ site.

This calculation can now be directly compared to the analysis of a high statistics ZF-$\mu$SR measurement performed at 5~K on a single crystal 
obtained according to the same procedure as the one used for the TF-$\mu$SR measurements.  
We note first that the oscillatory part of the muon polarization function associated with the field distribution given by Eq.~(\ref{eq_fd}), and given by 
\begin{equation}
P_{{\rm osc},i}(t) = \int\displaylimits_{B_{{\rm min},i}}^{B_{{\rm max},i}}D_i(B_{\text{loc,\tiny ZF}})\,\cos(\gamma_\mu B_{\text{loc,\tiny ZF}}\,t)~,
\label{p_osc}
\end{equation}
cannot be obtained analytically. We therefore approximate $D_i(B_{\text{loc,\tiny ZF}})$ with a shifted Overhauser distribution
\begin{figure}[b]
\includegraphics[width=0.9\columnwidth]{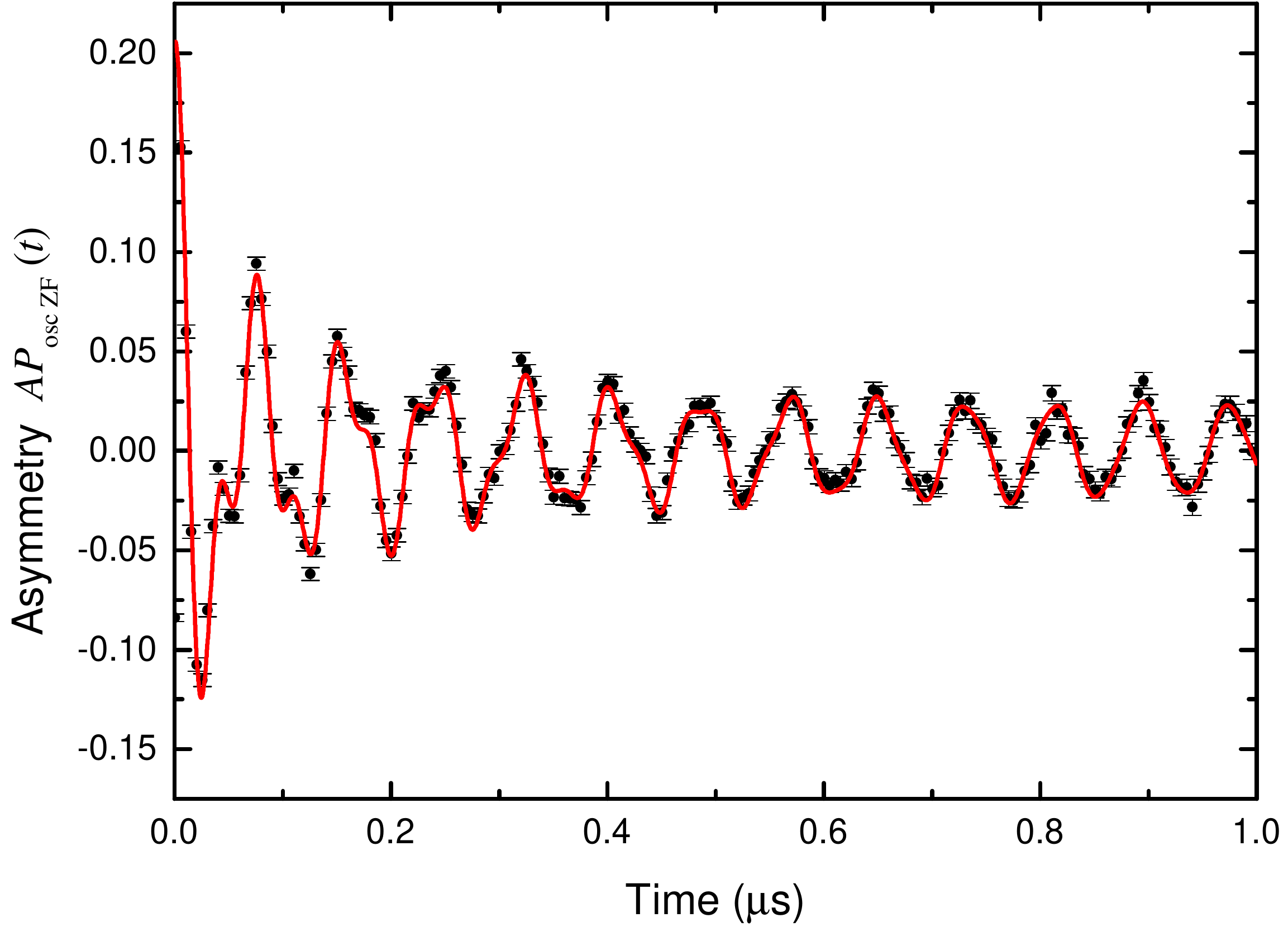}
\caption{(Color online) ZF-$\mu$SR data taken at 5~K (initial muon polarization parallel to the [111] direction. The line represents a fit of Eq.~(\ref{pol_func}) to the data using the parameters given in the text (for a better visibility a time-binning corresponding to 5~ns was choosen, whereas the fits were performed with a time-binning of 1.25~ns).} \label{fig_ZF_muSR_spectra_5K}
\end{figure}
\begin{equation}
D_i(B_{\text{loc,\tiny ZF}})\simeq\frac{1}{\pi}\frac{1}{\sqrt{\Delta B_i^2-(B_{\text{loc,\tiny ZF}}-B_{\text{av},i})^2}}~,
\label{eq_fd_approx}
\end{equation}
where $\Delta B_i=(B_{{\rm max},i}-B_{{\rm min},i})/2$ and $B_{\text{av},i}=(B_{{\rm max},i}+B_{{\rm min},i})/2$. The upper panel of Fig.~\ref{fig_approx_Bessel} shows the difference between the exact and approximated field distribution, which is symmetrical with respect to the singularities, i.e. some weight is transfered from the upper cutoff to the lower cutoff field. The lower panel of Fig.~\ref{fig_approx_Bessel} shows the oscillatory part of the muon polarization function obtained for both field distributions. For the exact field distribution a numerical calculation was performed using Eq.~(\ref{p_osc}), whereas for the approximated field distribution the muon polarization function can be obtained analytically and is given by
\begin{equation}
P_{{\rm osc\, Overh},i}\,(t) = J_0(\gamma_\mu \Delta B_i\,t)\,\cos(\gamma_\mu B_{\text{av},i}\,t)~,
\label{p_osc_approx}
\end{equation}
where $J_0$ is a Bessel function of the first kind. We see that the function $P_{{\rm osc\, Overh},i}\,(t)$ catches the essential features of $P_{{\rm osc},i}(t)$ and constitutes a good approximation. Having this in mind, we can now write a function which will be fitted to the ZF experimental data, i.e.
\begin{figure}[t]
\includegraphics[width=0.8\columnwidth]{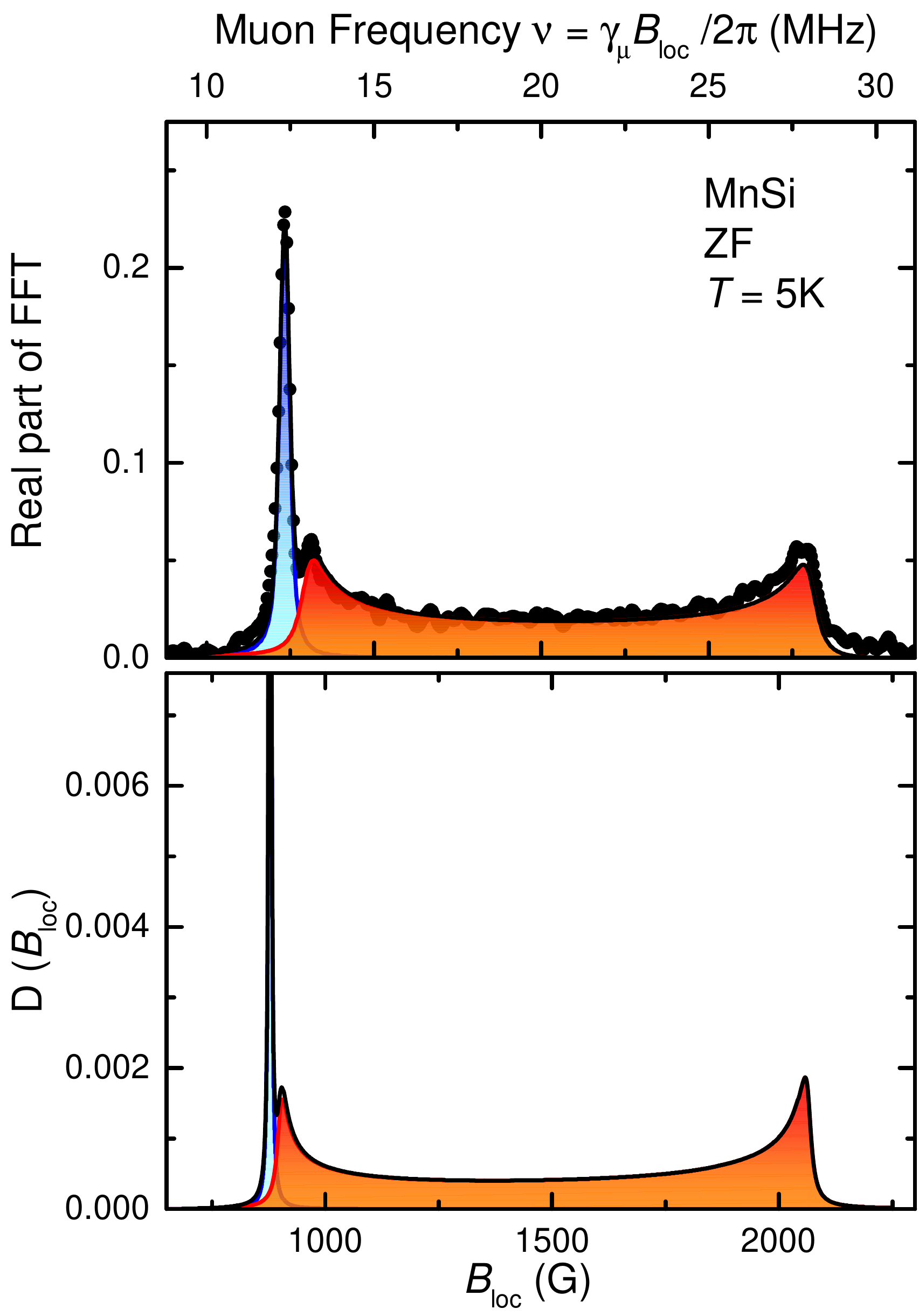}
\caption{(Color online) Upper panel: Fast Fourier Transform of the ZF-$\mu$SR data obtained at 5~K and shown on Fig.~\ref{fig_ZF_muSR_spectra_5K}. The blue and red components represent respectively the first and second component of the right-hand side of Eq~(\ref{pol_func}). Lower panel: computed field distributions for the $I$-site (blue) and for the $II$, $III$ and $IV$ sites (red). For a better visibility, these field distributions have been slightly folded by Lorentzian distributions with widths (FWHM) of 5 and 20~G, respectively. } \label{fig_simul_ZF}
\end{figure}
\begin{alignat}{4}
\mathcal{A}\,P_{{\rm osc\,\tiny ZF}}(t)&=\,&&\mathcal{A}_1\cos(\gamma_\mu B_1\,t+\psi_1)\,\exp(-\lambda_1t)\nonumber\\
&\,&&+\mathcal{A}_2J_0(\gamma_\mu \Delta B_2\,t)\cos(\gamma_\mu B_{\text{av},2}\,t+\psi_2)\nonumber \\
&\,&&\qquad\times \exp(-\lambda_2t)~,
\label{pol_func}
\end{alignat}
where the first component is associated to the site $I$ and the second to the sites $II$, $III$ and $IV$. The depolarization rates reflect as usual any static or dynamical effects, the discussion of which goes beyond the scope of this study. Figure~\ref{fig_ZF_muSR_spectra_5K} shows the ZF-$\mu$SR spectra with the fitted polarization function. From the fitted parameters one obtains $B_{\text{{\tiny\it I}}}^{\text{exp}}= 911(2)$~G (12.34~MHz), $B_{\text{min,\tiny\it II,III,IV}}^{\text{exp}}= 959(3)$~G (12.99~MHz) and $B_{\text{max,{\tiny\it II,III,IV}}}^{\text{exp}}= 2071(3)$~G (28.06~MHz). These values agree very well with our calculations performed without free parameters. In addition, both phase parameters $\psi_1$ and $\psi_2$ were found to be compatible with zero, as expected. The depolarization rates are found to have the values $\lambda_1=0.70(03)\,\mu{\rm s}^{-1}$ and $\lambda_2=1.78(16)\,\mu{\rm s}^{-1}$. We note that the small difference between the values of $B_{\text{{\tiny\it I}}}^{\text{exp}}$ and $B_{\text{min,\tiny\it II,III,IV}}^{\text{exp}}$ (which is also reflected in our calculations) perfectly explains that solely two peaks were invoked in former $\mu$SR data realized with much less statistics. Finally an interesting point is the fact that  $B_{\text{{\tiny\it I}}}<B_{\text{min,\tiny\it II,III,IV}}$ in both the calculations and the experimental data. This is remarkable as the position of $B_{\text{min,\tiny\it II,III,IV}}$, with respect to $B_{\text{{\tiny\it I}}}$, depends on the helicity of the incommensurate helix and is located above $B_{\text{{\tiny\it I}}}$ only for left-handed helicity. It is therefore tempting to take the ZF-$\mu$SR data as a late confirmation of the helix helicity. We note also that a close look at the Fast Fourier Transform reveals as expected a very slight divergence near the cutoff fields between the data and the shifted Overhauser distribution (see the upper panel of Fig.~\ref{fig_simul_ZF} and compare the data points to the red distribution). We therefore conclude here that from the knowledge of the muon site determined by our TF-data, one can utterly explain the characteristics of the ZF-$\mu$SR data.   

\subsection{Muon Site: Comparing Experimental Determination and Ab-Initio Calculations}
For completeness, we compare now our precise experimental determination of the muon site with {\it ab initio} calculations based on the density-functional theory (DFT). Such calculations have recently been shown to accurately reproduce the muon sites observed in different materials as wide-gap semiconductors, insulating systems or cuprates and iron-based high-$T_c$ superconductors \cite{DFT1,DFT2,DFT3,DFT4}. The calculations were performed at the University of Parma and we like to stress that they have been performed in a {\it blind modus}, i.e. without any prior knowledge of the muon site deduced from the TF-$\mu$SR data.

\begin{figure}[t]
\includegraphics[width=0.85\columnwidth]{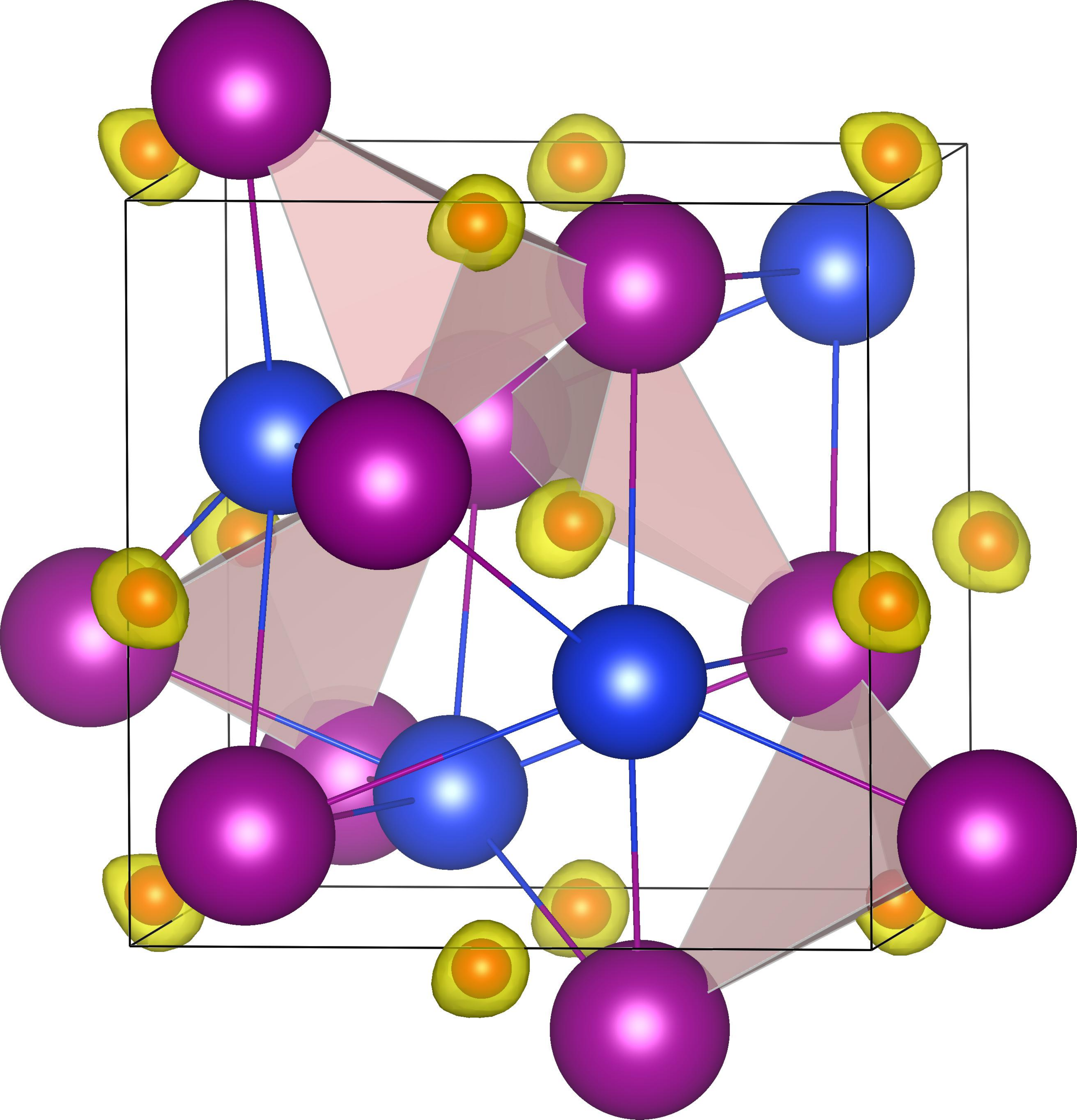}
\caption{(Color online) 
Sketch of the crystallographic structure of MnSi with minima regions of the electrostatic
potential (yellow). The rest of the color scheme is the same as in Fig. \ref{Structure}. 
The yellow regions define isosurfaces with an energy set to $V = E_0/2$ with $E_0$ being the ground state energy
obtained from the solution of the Schr\"odinger equation for the muon in the
electrostatic potential. The equivalent muon stopping sites determined by TF-$\mu$SR are also indicated and are enclosed in the minima regions. Note that compared to Fig. \ref{Structure}, we also report the muon sites located just outside of the primary unit cell, which are enclosed in minima regions extending into the primary unit cell.
} \label{fig_DFT}
\end{figure}
To analyze the electronic structure of MnSi we use the FP-LAPW approach as implemented in the \textsc{Elk} code \cite{elk}.
The generalized gradient approximation (GGA) as formulated by Perdew, Burke and Ernzerhof is used to approximate the exchange and correlation potential\cite{Perdew}. The cutoff for the plane wave expansion in the interstitial region is set to $|\mathbf{G + k}|_{\mathrm{max}} \cdot R_{\mathrm{mt}}$ = 9 ($R_{\mathrm{mt}}$ being the smallest muffin-tin radius). The reciprocal space is sampled on a $16\times16\times16$ Monkhorst-Pack\cite{Monkhorst} grid.
The experimental atomic positions and lattice constant are used. A ferromagnetic ground state is considered and the resulting magnetic moment on the Mn atoms is approximately 1~$\mu_{B}$. Indeed MnSi in a well known case in which the mean field approximation leads to a overestimation of the magnetic moments \cite{Jeong,Collyer}.

The muon site is identified from the minima of the electrostatic potential obtained from the ground state electronic density of the unperturbed material. There are four equivalent minima in the unit cell, as show in Fig.~\ref{fig_DFT}, which correspond to the $4a$ Wyckoff position with fractional coordinates 
(0.523,0.523,0.523). This coarse estimation nicely agrees with the experimentally evaluated site.

The standard procedure\cite{DFT1,DFT3} for the DFT site assignment further requires the analysis of the perturbation introduced by the muon in the vicinity of its embedding site. This is done by analyzing the electronic and crystallographic modifications introduced by the charged impurity. The muon site is eventually validated by considering the spread of its wave-function.
These steps are computationally demanding and outside the scope of the present work. However preliminary indications suggest that the inclusion of the muon in a supercell produces an even closer agreement with the experimental value, providing a 4$a$ muon stopping site at the coordinates (0.538,0.538,0.538).

\section{Conclusions}
We conclude that the observed angular dependence of the TF-$\mu$SR signals (except for a very slight dependence due to the demagnetization factor) can be beyond any doubt ascribed by a muon sitting at a 4$a$ Wyckoff position and detecting the dipolar field produced by the Mn moments. Consequently  the $\mu$SR response of paramagnetic MnSi under the application of a magnetic field can be fully understood without invoking a hypothetical magnetic polaron state. 
In addition we have shown that the knowledge of the muon stopping site provides a clear understanding of the ZF-$\mu$SR data and that the computed field distribution perfectly agrees with the measured one. Moreover, {\it blind modus ab initio} DFT calculations provide a 4$a$ muon stopping which is in very close agreement with the site determined by the experimental $\mu$SR data. 

\begin{acknowledgments}
D.A. acknowledges partial financial support from the Romanian UEFISCDI Project No. PN-II-ID-PCE-2011-3-0583 (85/2011). R.D.R., P.B. and F.B. acknowledge partial support of PRIN grant 2012X3YFZ2\_004.
Part of this work was performed at the Swiss Muon Source, Paul Scherrer Institut, Villigen, Switzerland. 
\end{acknowledgments}


\begin{thebibliography}{99}
\bibitem{Amato_foot_1} Note that in Ref.~\onlinecite{YaouancDalmas} and \onlinecite{Feyerherm} the Eq.~(\ref{Adipgeneral}) is expressed with the terms containing the polynomial $P_2^1$ as being  positive. The difference arises from the definition of the associated Legendre polynomial. We adopt here the common expression, i.e. $P_2^1(\cos \theta')=-3\sin \theta'\cos \theta'$~.
\bibitem{Amato_foot_2} Bear in mind that, strictly speaking, 
we should actually only consider the contribution of the $3d$ electrons to the total susceptibility. However, the fact that the total susceptibility perfectly follows a $1/T$-temperature dependence \cite{suscep1,suscep2,suscep3} is a strong indication that the response of the local moments will dominate the susceptibility upon decreasing the temperature.  As our measurements were performed at rather low temperature we can safely assume that $\chi \simeq \chi_{3d}$. In any case, as explained later in the text, an even large uncertainty on the value of the susceptibility will only very weakly affect the position of the obtained muon site. However, the conclusion that the muon is
stopped at a $4a$ position is not affected by a variation of $\chi$.
\bibitem{Amato_foot_3}Even though the wavelength of the magnetic structure is long, the system is antiferromagnetic and the Lorentz field is zero, since the magnetization integrated outside of the Lorentz sphere vanishes.
\end{thebibliography}
\end{document}